\documentclass{iopart}
\usepackage{graphicx}

\begin{document}

\title[PHENIX High-$p_{\rm T}$ hadron and photon production]{Systematic study of high-$p_{\rm T}$ hadron and photon production with the PHENIX experiment}

\author{Christian Klein-B\"osing$^1$ for the PHENIX Collaboration}
% \footnote[3]{For the full PHENIX Collaboration author list and
% acknowledgments, see Appendix "Collaborations" of this volume.}  }

\address{$^1$CERN, CH-1211 Geneva 23, Switzerland}

\ead{Christian.Klein-Boesing@cern.ch}
\begin{abstract}

The suppression of hadrons with large transverse momentum ($p_{\rm T}$) in
central Au+Au collisions at $\sqrt{s_{\rm NN}}$ = 200 GeV compared to a
binary scaled p+p reference is one of the major discoveries at
RHIC. To understand the nature of this suppression PHENIX has
performed detailed studies of the energy and system-size dependence of
the suppression pattern, including the first RHIC measurement near SPS
energies.  An additional source of information is provided by direct
photons.  Since they escape the medium basically unaffected they can
provide a high $p_{\rm T}$ baseline for hard-scattering processes.

An overview of hadron production at high $p_{\rm T}$ in different colliding
systems and at energies from $\sqrt{s_{\rm NN}} = 22.4 - 200$~GeV will be
given.  In addition, the latest direct photon measurements by the
PHENIX experiment shall be discussed.
\end{abstract}

\pacs{25.75.Bh, 25.75.Cj}


% \maketitle

\section{Introduction}

One of the primary goals of the PHENIX experiment at the Relativistic
Heavy Ion Collider (RHIC) at Brookhaven National Laboratory is to
study strongly interacting matter under extreme conditions. In
particular the creation of a new state of matter, the Quark-Gluon
Plasma (QGP) which is expected to form when energy densities above
$\epsilon_0 \approx 1$~GeV/fm$^3$ are attained in the collision.

One proposed signature of this new phase is the suppression of hadrons
with large transverse momentum ($p_{\rm T}$) compared to the expectation
from scaled p+p reactions \cite{Gyu90a,Gyu94b}.  The production of
these particles is dominated by the scattering of partons with large
momentum transfer $Q^2$ ($\propto p_{\rm T}^2$), so-called \emph{hard}
scattering, and their subsequent fragmentation into observable
particles in \emph{jets} along the original parton direction.

\begin{figure}[!t]
   \centerline{\includegraphics[width=\linewidth]{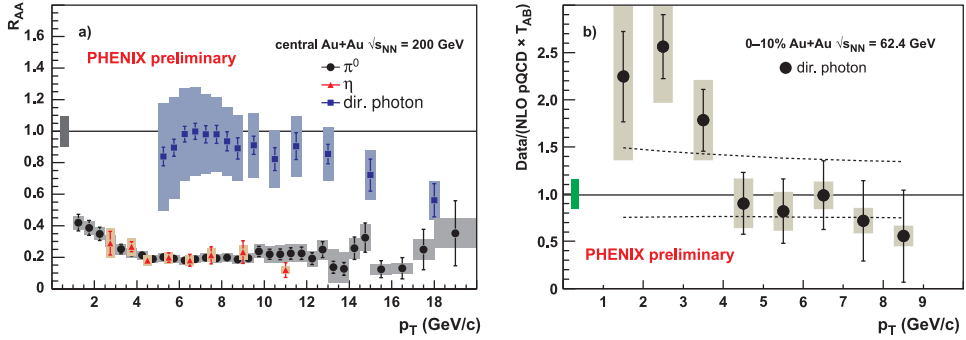}}
   \caption{a) Nuclear modification factor for $\pi^0$s, $\eta$s and
   direct photons in central Au+Au collisions at $\sqrt{s_{\rm NN}} =$
   200 GeV. $p_{\rm T}$ uncorrelated uncertainties are given as error
   bars, correlated as boxes around the points. The normalization
   uncertainty is shown by the box at unity. b) $R_{AA}$ for direct
   photons in central Au+Au collisions at $\sqrt{s_{\rm NN}}
   =$~62.4~GeV, uncertainties as in a), the shift with respect to
   unity when using different scales ($\mu =$ 0.5, 2.0 $p_{\rm T}$) in
   the pQCD reference calculation is given by the dashed lines.}
   \label{fig1}
\end{figure}

The first step to quantify medium effects on particle production is to
compare single particle spectra in heavy ion collisions to the
expectation for the QCD vacuum.  This is usually done by the nuclear
modification factor $R_{AA}$, which compares the particle production
in $A$+$A$ collisions to elementary p+p collisions scaled by a
geometrical factor, the nuclear overlap function $T_{AA}$:
\begin{equation}
R_{AA} = \frac{\rmd^2N_{AA}/ \rmd y \rmd p_{\rm T}} {T_{AA} \cdot \rmd^2
\sigma_{\rm pp}/ \rmd y \rmd p_{\rm T}}.
\end{equation}
$T_{AA}$ accounts for the increased number of nucleons in the incoming
$A$-nuclei and is related to the number of binary collions $N_{\rm
coll}$ by $T_{AA} \approx N_{\rm coll}/\sigma_{\rm inel}^{\rm pp}$. In
the absence of any medium effects and at sufficiently high $p_{\rm
T}$, where hard scattering is the dominant source of particle
production $R_{AA}$ should be unity, any deviation from unity
indicating the influence of the medium.

Indeed, already the first hadron measurements in Au+Au reactions at
RHIC showed a hadron yield suppressed up to a factor of five in
central collisions \cite{Adc02a,Adl03d,Ada03b}. This suppression can
be explained by the energy loss of hard scattered partons via induced
gluon bremsstrahlung in a medium with high color density.  However,
with the first hadron measurements it was not possible to distinguish
experimentally between final state effects, e.g. parton energy loss,
and the effects of cold nuclear matter. These initial state effects
can enhance or suppress the particle production compared to p+p and
involve e.g. multiple soft scattering of incoming partons or nucleons
(Cronin effect), or a modification of the parton distribution function
in a nucleus (shadowing, anti-shadowing or gluon-saturation).

Initial state effects have subsequently been ruled out as a source of
the observed suppression by two key results in d+Au and Au+Au at
$\sqrt{s_{\rm NN}} = 200$~GeV:
\begin{itemize}
\item Charged hadrons and neutral pions are not suppressed in d+Au
collisions (cold nuclear matter) \cite{Adler:2003ii}.
\item The nuclear modification factor for direct photons at high $p_{\rm T}$
is consistent with unity \cite{Adler:2005ig}.
\end{itemize}
 
The latter provides the in-situ control for parton energy loss and for
the assumption of a geometrical scaling for hard scattering, since
direct photons are also produced in initial hard scattering but are
not affected by the strong interaction.  They can traverse the plasma
basically unaltered and at high $p_{\rm T}$ the direct photon yield should
directly reflect the initial state of quarks and gluons in the
nucleus. Going to lower $p_{\rm T}$ the measurement of direct photons in
$A$+$A$ collisions can provide further information on the medium, since
other photon sources become important. These involve e.g. the
interaction between hard scattered and thermal partons,
photon bremsstrahlung in the QGP and thermal radiation.

Since the first observation of suppressed particle production the
study of high $p_{\rm T}$ particles via the nuclear modification factor has
been a major tool for studying the dense nuclear matter. In fact, the
observations at RHIC triggered also an extensive search for similar
effects in data taken at the CERN/SPS (see e.g. \cite{d'Enterria:2004ig,Aggarwal:2007gw}). 

In this paper we will concentrate on the most recent measurements of
identified particles by the PHENIX experiment at high $p_{\rm T}$ for
different colliding systems and energies. The main information in this
region is provided by the measurement of neutral mesons
($\pi^0$,$\eta$) and direct photons, this allows to study different
mesons from parton fragmentation and direct photons produced in early
hard scatterings.

\section{Data analysis and results}

\begin{figure}[!t]
   \centerline{\includegraphics[width=\linewidth]{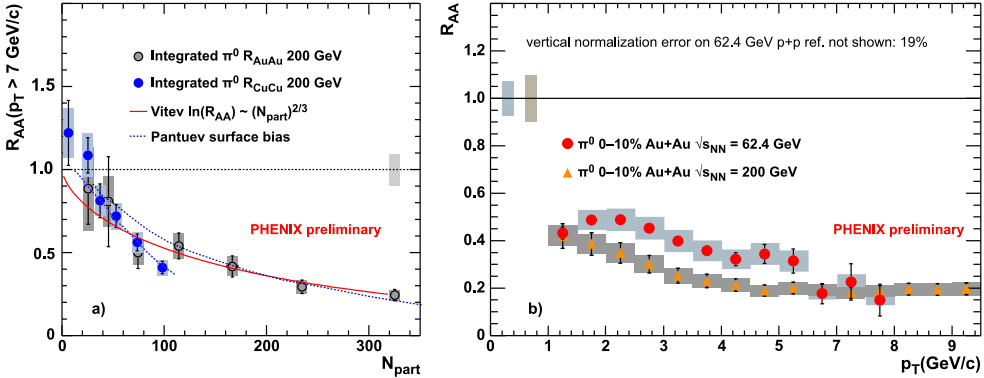}}
   \caption{a) Integrated nuclear modification factor for Au+Au and
   Cu+Cu collisions compared to a scaling with $N_{\rm part}^{2/3}$
   \cite{Vitev:2005jx} and a model with surface/volume effects
   \cite{Pantuev:2005jt}. b) Nuclear modification factor in central
   Au+Au collisions at $\sqrt{s_{\rm NN}}$ = 200 and 62.4
   GeV. Uncertainties as described in Figure~\ref{fig1}, the
   19\% normalization error of the p+p data  at 62.4 GeV is not shown.
   \label{fig2}}
\end{figure}

The data presented here has been collected by the PHENIX experiment
during three different RHIC runs and spans the following collision
species and energies:
\begin{center}
\begin{tabular}[!h]{|c|c|l|}
\hline
RHIC run  & Species & $\sqrt{s_{\rm NN}}$~(GeV) \\
\hline\hline
4 & Au+Au & 200, 62.4 \\
\hline
5 & Cu+Cu & 200, 62.4, 22.4\\
  & p+p   & 200\\
\hline
6 & p+p   & 200, 62.4\\
\hline
\end{tabular}
\end{center}
The choice of copper as colliding system has been motivated by the
fact, that the suppression in Au+Au sets in at a centrality
corresponding to approximately $50-100$ participating nucleons
($N_{\rm part}$), see Figure~\ref{fig2}a). Copper allows to
study this range in more detail at a more spherical collision
geometry.

All measurements have been performed using the electromagnetic
calorimeter EMCal. It is the outermost detector of the PHENIX central
arm and provides a solid angle coverage of $-0.35 < \eta < 0.35$ and
$\Delta\phi = \pi$. The $\pi^0$ and $\eta$ mesons are reconstructed
via their two-photon decay with an invariant mass analysis of photon
pairs. The combinatorial background is determined by a \emph{mixed
event} technique, where photon pairs from different events are
combined to form the uncorrelated background.  The raw spectra are
corrected for the geometrical acceptance and for reconstruction
efficiency losses.

Direct photons are by definition all photons not originating from
radiative decays. The direct photons can thus be determined by
comparing the inclusive photon spectrum, obtained after subtraction of
charged particle and neutral hadron contributions to the EMCal cluster
spectrum and correction for efficiency and acceptance losses, to the
expectation from decay photons, which are mainly originating from
$\pi^0,\eta\rightarrow \gamma\gamma$. Any excess above this
expectation is considered as a direct photon signal \cite{Adler:2005ig}.

\section{Au+Au Collisions at $\sqrt{s_{\rm NN}}$ = 200 GeV}

The factor of 15 more sampled events in Au+Au collisions during RHIC's
fourth run compared to the first data taking at design energy allows
to study the production of neutral pions up to $ p_{\rm T} = 20$~GeV/$c$. A
similar improvement in the $p_{\rm T}$ range was also reached in the p+p
reference data. The resulting nuclear modification factor shown in
Figure~\ref{fig1}a) is basically flat up to the highest $p_{\rm T}$.

The fact that neutral pions and $\eta$s show the same amount of
suppression indicates that the jet-quenching mechanism does not depend
on the mass of the (light-quark) meson. This is expected when the
suppression only depends on the energy loss of the parent quark
and not on the leading hadron itself, which is produced in a universal
fragmentation process.

The direct photon result confirms the measurement from
\cite{Adler:2005ig}, a nuclear modification factor of unity up to $p_{\rm T}
= 14$~GeV/$c$, which can be attributed to the insensitivity of hard
scattered photons to the produced medium. However, at higher $p_{\rm T}$ a
decrease of the direct photon yield becomes apparent. One may
speculate on the influence of photons from parton fragmentation, which
would be suppressed in Au+Au, but this effect should rather decrease
with $p_{\rm T}$. A simpler explanation can be given by the influence of
protons and neutrons in the nucleus.

When studying strong processes the difference between n and p is
basically negligible. This is not the case in the production of direct
photons, where the scattering (e.g. $qg \rightarrow q\gamma$) involves
an electromagnetic vertex. However, the normalization to p+p in the
nuclear modification factor does not account for the fact that already
the elementary processes p+p, p+n, and n+n are different when the
valence quark distributions become important at a certain probed
momentum fraction $x_T \approx p_{\rm T}/\sqrt{s_{\rm NN}}$ . This trivial
isospin effect can be estimated via a pQCD calculation of the
elementary processes to be of the order of 15\% for $x_T = 0.1$
\cite{Sakaguchi:2007zs}. However the interplay between different
effects, isospin, modified structure function and the influence of
fragmentation photons should also be considered (see
e.g. \cite{Arleo:2006xb}). 

An experimental handle on the isospin effect is given by the study of
direct photons at lower collision energy, which allows to study the
effect at the same $x_T$ but lower $p_{\rm T}$. The first results on
direct photon production at $\sqrt{s_{\rm NN}} = 62.4$~GeV shown in
Figure~\ref{fig1}b) indicate a similar suppression at the expected
lower $p_{\rm T}$, but the statistical uncertainties as well as the
missing p+p reference at this energy prevent any firm conclusions at
the moment.

\section{System size and energy dependence}

 The study of different system sizes is motivated by the fact that
smaller systems allow a better discrimination for smaller values of
$N_{\rm part}$ and hence at a smaller energy density. E.g. Cu+Cu ($A =
63$) collisions allow to sample the $N_{\rm part}$-range of peripheral
Au+Au ($A = 197$) collisions. In addition, it provides an insight into
the influence of the collision geometry, since at the same $N_{\rm
part}$ the overlap region is more spherical for smaller systems.

It is found that the $p_{\rm T}$ dependence of the nuclear modification
factor at the same $N_{\rm part}$ is very similar for Au+Au and Cu+Cu
collisions (not shown, see e.g. \cite{KleinBoesing:2006kd}). This is
also seen in the centrality dependence of the integrated nuclear
modification factor for Cu+Cu and Au+Au in
Figure~\ref{fig2}a). $R_{AA}$ follows approximately a scaling with
$\ln(N_{\rm part}^{2/3})$, which is inspired by the volume dependence of
the energy density \cite{Vitev:2005jx}. However, the slightly
different slopes can be better explained when taking into account
surface effects as e.g. in \cite{Pantuev:2005jt}, where it is assumed
that only jets originating from a certain depth below the surface can
be observed.

A important constrained to any description of the in-medium energy
loss is provided by the measurement of the energy dependence. Since at
different energies not only the energy density varies but also the
influence of initial state effects, such as the Cronin enhancement and
shadowing, changes as well as the slope of the spectra.

For the determination of the nuclear modification factor the knowledge
of the baseline p+p reference is crucial. This is nicely illustrated
by the $R_{AA}$ for central Au+Au at $\sqrt{s_{\rm NN}} = 62.4$~GeV shown
in Figure~\ref{fig1}b) and constructed with the measured PHENIX p+p
reference. Compared to previous comparisons which only employed a
reference fit to ISR data, it shows that already at 62.4 GeV the
suppression in $R_{AA}$ is more than a factor of two and follows a
similar $p_{\rm T}$ dependence as the data at $\sqrt{s_{\rm NN}} = 200$~GeV.

\begin{figure}[!t]
   \centerline{\includegraphics[width=0.6\linewidth]{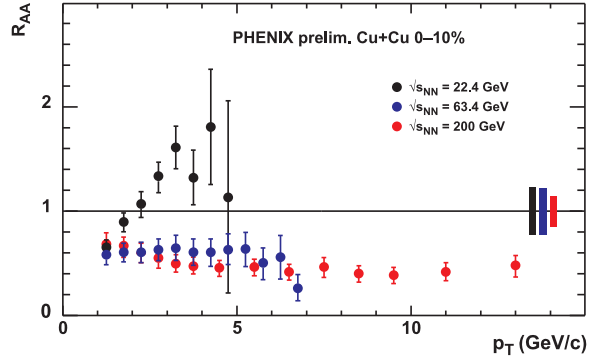}}
   \caption{Energy dependence of the nuclear modification factor in
   central Cu+Cu collisions. The combined normalization error due to
   the uncertainty in $T_{AA}$ and the scale uncertainty of the p+p
   reference is indicated by the error boxes around unity, systematic
   and statistical uncertainties are summed in quadrature and given by
   the error bars.}
   \label{fig3}
\end{figure}

This behavior is also seen in Cu+Cu collisions at $\sqrt{s_{\rm NN}} =
200$~GeV and $\sqrt{s_{\rm NN}} = 62.4$~GeV shown in
Figure~\ref{fig3}, again both energies show a very similar $p_{\rm T}$
dependence and magnitude of the nuclear modification factor. However,
one also has to consider that the slope of the scattered partons at
$\sqrt{s_{\rm NN}} = 62.4$~GeV is steeper compared to 200 GeV, so an
energy loss $\Delta E/E$ is more pronounced. Hence, the same $R_{AA}$
at different energies is not tantamount to the same parton energy
loss.

The nuclear modification factor close to CERN/SPS energies has been
measured for the first time at RHIC by the PHENIX experiment in Cu+Cu
collisions at $\sqrt{s_{\rm NN}} = 22.4$~GeV. Though the measurement of
p+p reactions at this energy within PHENIX is still missing, the
wealth of data for p+p $\rightarrow \pi^0/\pi^{\pm}$+$X$ in the range
of $\sqrt{s_{\rm NN}} = 21.7 - 23$~GeV allows to construct a reference
with a systematic uncertainty of $\approx 20\%$. The resulting
$R_{AA}$ is also shown in Figure~\ref{fig3} and exhibits no
suppression, instead it indicates an enhancement.  To understand
whether parton energy loss is present but compensated by
Cronin enhancement the measurement of Au+Au collisions is needed to
reach higher energy densities, as well a solid p+p reference at the
same energy.

\section{Conclusion}

We have shown a systematic study of neutral meson and direct photon
production at high $p_{\rm T}$ measured with the PHENIX experiment, based
on the nuclear modification factor for different particle species,
colliding systems and center-of-mass energies.

The nuclear modification factor in central Au+Au at top RHIC energies
is flat up to $p_{\rm T} = 20$~GeV/$c$ and shows a similar suppression
for $\pi^0$s and $\eta$s while for direct photons no suppression is
observed over a wide $p_{\rm T}$ range. The deficit of direct photons at the
highest $p_{\rm T}$ may be due to an isospin effect, but the interplay
between different nuclear and medium effects needs to be investigated
further.

The nuclear modification factor is similar in the two colliding
systems Cu+Cu and Au+Au at a similar $N_{\rm part}$, with indication
that geometry effects need to be considered. We have also shown the
first study of the energy dependence of $R_{AA}$ from $\sqrt{s_{\rm NN}}
\approx 20 - 200$~GeV, within the same experiment. It shows that
suppression already is a dominating effect at $\sqrt{s_{\rm NN}} =
62.4$~GeV, while close to CERN/SPS energies an enhancement is
observed.

\end{document}